\documentclass[prl,twocolumn,showpacs,amsmath,amssymb,superscriptaddress]{revtex4-1}
\usepackage{graphicx}% Include figure files
\usepackage{bm}
\usepackage{bbm}
\usepackage{ulem}
\usepackage{color}

\newcommand{\nc}{\newcommand}

\nc{\be}{\begin{equation}} \nc{\ee}{\end{equation}}
\nc{\bea}{\begin{eqnarray}} \nc{\eea}{\end{eqnarray}}
\nc{\bean}{\begin{eqnarray*}} \nc{\eean}{\end{eqnarray*}}

\begin{document}

\title{Reading charge transport from spin dynamics on the surface of topological insulator}
\author{Xin Liu}
\affiliation{ Department of Physics, Texas A\&M University, College
Station, TX 77843-4242, USA}
\affiliation{ Department of Physics, The Pennsylvania State University, University Park, PA 16802-6300, USA}
\author{Jairo Sinova}
\affiliation{ Department of Physics, Texas A\&M University, College
Station, TX 77843-4242, USA}
\affiliation{Institute of Physics ASCR, Cukrovarnick\'a 10, 162 53 Praha 6, Czech
Republic}
\date{\today}

\begin{abstract}
Resolving the conductance of the topological surface states (TSSs) from the bulk contribution has been a great challenge for studying the transport property of topological insulators. By developing a non-purturbative diffusion equation that describes fully the spin-charge dynamics in the strong spin-orbit coupling regime, we present a proposal to read the charge transport information of TSSs from its spin dynamics which can be isolated from the bulk contribution by time-resolved second harmonic generation pump-probe measurement.
We demonstrate the qualitatively different Dyaknov-Perel spin relaxation behavior between the TSSs and the two-dimensional spin-orbit coupling electron gas. The decay time of both in-plane and out-of-plane spin polarization is naturally proved to be identical to the charge transport time. The out-of-plane spin dynamics is shown to be in the experimentally reachable regime of the femtosecond pump probe spectroscopy and thereby we suggest experiments to detect the charge transport property of the TSSs from their unique spin dynamics.
\end{abstract}

\maketitle

Topological insulators (TIs) are a class of time-reversal invariant matter \cite{Hasan:2010_a,Qi:2011_a}.
Recently, both theory \cite{Burkov:2010_a,Culcer:2010_a,Gao:2011_a} and experiments \cite{Qu:2010_a, Analytis:2010_a, Hsieh:2011_a, Hsieh:2011_b, WangXiaolin:2012_a,Zhang:2012_a,Li:2012_a} focus on the transport property of the TIs which have spin-momentum locking surface states. The helical topological surface states (TSSs) break the Fermi doubling and have very strong spin-orbit coupling (SOC) which gives their unique charge and spin transport properties \cite{Raghu:2010_a,Tse:2010_a}. However,
in most TIs the bulk states also contribute the conductance, which makes it a great challenge to extract the transport properties of the TSSs, such as mobility, from the experimental data \cite{Qu:2010_a, Analytis:2010_a}. On the other hand, the spin polarization of the TSSs has been detected by the time-resolved second harmonic generation pump-probe measurement which can separate the surface response from the bulk \cite{Hsieh:2011_a,Hsieh:2011_b}. However the complete spin dynamical theory of TSS is still absent due to the strong SOC of TSSs which can not be treated perturbatively as it is usually done in the traditional diffusive equation. Here the very strong SOC means $\Omega_{\rm so}\tau_{\rm p} \gg 1$ where $\Omega_{\rm so}$ is the spin precession frequency due to the SOC and $\tau_{\rm p}$ is the momentum scattering time.

In this work, we developed a non-perturbative spin dynamic theory to fully capture the spin dynamics of the TSSs which reveal the qualitative difference to that in other SOC 2DEGs such as GaAs quantum well \cite{Burkov:2004_a,Bernevig:2006_a,Stanescu:2007_a,LiuXin:2011_a}. The spin polarization along x,y,z direction are decoupled to each other even though the TSSs have strong SOC. This is much different to the spin dynamics in the 2DEGs where the different spin components are coupled for finite spin wave length in both weak and strong SOC regime and the coupling strength is proportional to the SOC \cite{Burkov:2004_a,Burkov:2004_a,Bernevig:2006_a,Stanescu:2007_a,LiuXin:2011_a}. The decay time of both in-plane and out-of-plane spin polarization is predicted to be exactly equal to the charge transport time, $\tau_{\rm{tr}}$, instead of being proportional to the inverse of $\Omega_{\rm so}^2\tau_{\rm p}$, which is commonly the label of Dyakonov-Perel (DP) mechanism. We estimate the time evolution of the out-of-plane spin dynamics which we show to be in the experimentally reachable regime of the femtosecond pump probe spectroscopy \cite{Hsieh:2011_b}. Because the
surface spin dynamics can be isolated from the bulk contribution by  recent second harmonic
pump probe measurement \cite{Hsieh:2011_a,Hsieh:2011_b}, we propose to read charge transport property of TSSs from the spin by the pump probe measurement.

We assume that the low energy TSS has the Dirac-like Hamiltonian \cite{Hasan:2010_a,Qi:2011_a}
 \begin{eqnarray}\label{Ham-1}
 \hat{H}=v(\bm{z}\times \bm{\sigma})\cdot \mathbf{k}+H'(\bm{r}),
 \end{eqnarray}
 where $v$ is the constant velocity, $\bm{z}$ is the unit vector perpendicular to the surface of TI, $\bm{k}$ is the wave vector of the electron, $\sigma$ is pauli matrix and $H'(\bm{r})$ is the perturbative Hamiltonian which is dependent on the spacial coordinate $\bm{r}$. Here we assume $\hbar=1$ and consider only spin independent scattering. The velocity operator  has the form
 \begin{eqnarray}\label{vel-1}
 \hat{\bm{V}}=v{\bm{z} \times \bm{\sigma}},
 \end{eqnarray}
 and is proportional to the in-plane spin polarization perpendicular to the velocity direction. This special velocity operator makes the charge transport of the TSSs qualitatively different to the charge transport in the SOC 2DEG where the velocity operator is dominated by the momentum. As we know that the conductivity can be calculated from the current-current correlation function \cite{Mahan2010a} and the spin dynamics can be obtained by evaluate the spin-spin correlation function\cite{Burkov:2004_a} even in the presence of electron-electron interaction\cite{Punnoose:2006_a}. In the traditional 2DEG system, the velocity operator is proportional to the momentum and thereby the conductivity is actually calculated from the momentum-momentum correlation function\cite{Mahan2010a}. However for the TSSs, the current-current correlation function is proportional to spin-spin instead of momentum-momentum correlation function which make it possible to get the charge transport information from the spin dynamics. This statement should not be dependent on the special form of the perturbation Hamiltonian such as the disorder potential, electron-phonon and electron-electron interaction because the velocity operator is defined as $\hat{V}=-i[\bm{r},H]$ but $\bm{r}$ commutes with $H'(\bm{r})$. In current work, to simplify our discussion, we only focus on the non-magnetic short range impurity in the zero temperature.

The dynamics of the spin-charge polarization, as a non-equilibrium process, can be described by quantum Boltzmann equation \cite{Mishchenko:2004_a,LiuXin:2012_a}
 \begin{eqnarray}\label{GF-7}
\partial_t \hat{g}+\bm{\nabla_R}\cdot \{\frac{1}{2}\bm{\hat{V}},  \hat{g} \}+i[\hat{H}_0,\hat{g}]+\frac{\hat{g}}{\tau_{\rm p}}=\frac{\hat{\rho}}{\tau_{\rm p}}\,\,,
\end{eqnarray}
where $\hat{H}_0(\bm{k})=v(\bm{z}\times \bm{\sigma})\cdot \bm{k}$,
$\hat{g}(\theta,\bm{R},t)$ is the angular distribution function of the $2\times 2$ matrix of spin-$\frac{1}{2}$ Keldysh Green's function \cite{Mishchenko:2004_a,LiuXin:2012_a,Rammer:2007_a}, $\theta$ is the angle between the momentum and the $x$ axis, $\hat{\rho}(R,t)=\int \hat{g} d\theta/2\pi$
  is the density matrix and $\tau_{\rm p}$
 is the momentum scattering time at Fermi surface. %In this work, the operator with a hat means it is a matrix.
The difficulty of obtaining the spin dynamics equation in the strong SOC regime lies on the fact that the spin precession angle $2vk_{\rm f}\tau_{\rm p}$, is too large to be treated perturbatively. To circumvent this problem, we multiply $\sigma_j$ where
$j = 0,x,y,z$ on both sides of Eq.~\ref{GF-7} and calculate the trace. Using the fact that
$\rm{Tr}(\sigma_j \sigma_k)/2 = \delta_{jk}$,
the quantum Boltzmann equation can be written in the classical spin-charge 4D space after integrating out $E$ as \cite{LiuXin:2012_a,SM_T}
 \begin{eqnarray}\label{GF-9}
\hat{K}_{jk}g_{k} =i\rho_{j},
 \end{eqnarray}
 where
$g_j(\rho_{j})=\rm{Tr}[\hat{g}(\hat{\rho})\sigma_j/2]$. To simplify our discussion, we assume the uniform spin polarization along $y$ direction, say $\partial_y \hat{\rho}=0$. In this case, by multiplying the inverse of the matrix $\hat{K}$ on both sides of Eq.~\ref{GF-9} and integrating out the variable $\theta$, we obtain the non-purturbative spin-charge dynamic equation
 \begin{eqnarray}\label{GF-12}
 \rho_{j}=\hat{D}_{jk}\rho_k,
 \end{eqnarray}
where $\hat{D}=\int \frac{d\theta}{2\pi}\hat{K}^{-1}$ and $\hat{K}^{-1}$ has a complicated but analytically and numerically tractable form
\begin{widetext}
\begin{eqnarray}\label{GF-13}
\frac{\left(
\begin{array}{cccc}
 \Tilde{\Omega}  \left(\Tilde{\Omega} ^2+\Tilde{\Omega} _{\text{so}}^2\right) & i \cos (\theta ) \sin (\theta ) \tilde{\Delta} _x \Tilde{\Omega} _{\text{so}}^2 & -i \tilde{\Delta} _x \left(\Tilde{\Omega} ^2+\cos ^2(\theta ) \Tilde{\Omega} _{\text{so}}^2\right) & -i \Tilde{\Omega}  \sin (\theta ) \tilde{\Delta} _x \Tilde{\Omega} _{\text{so}} \\
 i \cos (\theta ) \sin (\theta ) \tilde{\Delta} _x \Tilde{\Omega} _{\text{so}}^2 & \Tilde{\Omega}  \left(\Tilde{\Omega} ^2+\tilde{\Delta} _x^2+\sin ^2(\theta ) \Tilde{\Omega} _{\text{so}}^2\right) & -\Tilde{\Omega}  \cos (\theta ) \sin (\theta ) \Tilde{\Omega} _{\text{so}}^2 & \cos (\theta ) \left(\Tilde{\Omega} ^2+\tilde{\Delta} _x^2\right) \Tilde{\Omega} _{\text{so}} \\
 -i \tilde{\Delta} _x \left(\Tilde{\Omega} ^2+\cos ^2(\theta ) \Tilde{\Omega} _{\text{so}}^2\right) & -\Tilde{\Omega}  \cos (\theta ) \sin (\theta ) \Tilde{\Omega} _{\text{so}}^2 & \Tilde{\Omega} ^3+\cos ^2(\theta ) \Tilde{\Omega} _{\text{so}}^2 \Tilde{\Omega}  & \Tilde{\Omega} ^2 \sin (\theta ) \Tilde{\Omega} _{\text{so}} \\
 i \Tilde{\Omega}  \sin (\theta ) \tilde{\Delta} _x \Tilde{\Omega} _{\text{so}} & -\cos (\theta ) \left(\Tilde{\Omega} ^2+\tilde{\Delta} _x^2\right) \Tilde{\Omega} _{\text{so}} & -\Tilde{\Omega} ^2 \sin (\theta ) \Tilde{\Omega} _{\text{so}} & \Tilde{\Omega}  \left(\Tilde{\Omega} ^2+\tilde{\Delta} _x^2\right)
\end{array}
\right)}{\left(\Tilde{\Omega} ^2+\Tilde{\Omega} _{\text{so}}^2\right) \Tilde{\Omega} ^2+\tilde{\Delta} _x^2 \left(\Tilde{\Omega} ^2+\cos ^2(\theta ) \Tilde{\Omega} _{\text{so}}^2\right)}
\end{eqnarray}
\end{widetext}

\begin{figure}
\centering
\begin{tabular}{l}
\includegraphics[width=0.6\columnwidth]{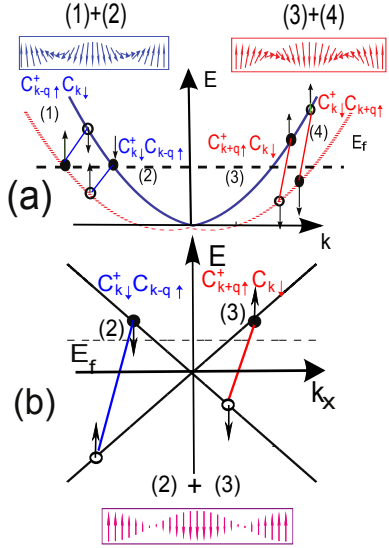}
\end{tabular}
\caption{ (a) Spin polarization configure at $k_y=0$ channel in 2DEG. The solid(hollow) circle means creating(annihilating) an electron into the system which is corresponding to $C^{\dagger}(C)$ operator. The clockwise SHM (the red arrows) in $x-z$ plane is constructed by $\sum_k c^{\dagger}_{k+q,\uparrow}c_{k,\downarrow}+\rm{h.c.}$ (the red creation and annihilation operators) and the counter-clockwise SHM (the blue arrows) is constructed by $\sum_k c^{\dagger}_{k-q,\uparrow}c_{k,\downarrow}+\rm{h.c.}$ (the blue creation and annihilation operators). Here $\uparrow(\downarrow)$ indicates spin along $+(-)$ y direction. (b) Spin polarization configure at $k_y=0$ channel. The collective spin polarization arises from the summation of the half clockwise SHM and half counter-clockwise.}
\label{Spinconfig}
\end{figure}

 Eq. (\ref{GF-12}) is a generalized spin diffusion equation and valid from the weak to the strong SOC regime. Because of the helical Hamiltonian, the denominator of the matrix $\hat{K}^{-1}$ is the function of $\cos^2\theta$ whose periodicity is $\pi$. By using the property of trigonometric functions without calculating the integral of $\theta$, we can easily prove that the spin polarizations along $x$, $y$ and $z$ direction are completely decoupled even though the TSSs have very strong SOC. This is very different to the spin wave of 2DEG where in the case of finite spin wave vector along $x$ direction, the SOC makes the spin polarization being rotated in $x-z$ plane shown in Fig~\ref{Spinconfig}a which is called spin helix(SH)\cite{Bernevig:2006_a}. On the TI surface, taking the spin polarization along $z$ direction for example, it is decoupled to any other spin components and therefore the SH is absent Fig~\ref{Spinconfig}b. This qualitative difference is because 2DEG respect the Fermi doubling but TSSs does not. For the spin polarization of the 2DEG shown in Fig.~\ref{Spinconfig}(a), there are two spin sub-bands on the Fermi surface which allow two spin helical modes(SHMs) with two different relaxation times\cite{Bernevig:2006_a,Weber:2007_a,Koralek:2009_a}. In Fig.~\ref{Spinconfig}(b), because the non-equilibrium state is around the Fermi surface and $E_{\rm f} \gg kT$, the spin sub-band below the Dirac point is still completely filled. Accordingly, at the $k_y=0$ channel the terms such as $\sum_{k>0}  C^{\dagger}_{k,\downarrow}C_{k+q,\uparrow}$ and $\sum_{k<0} C^{\dagger}_{k-q,\uparrow}C_{k,\downarrow}$, which add electrons to the spin sub-band below the Dirac point, will not contribute to spin polarization \cite{SM_T}. Here $\uparrow(\downarrow)$ indicates spin along $+(-)$ y direction. Therefore, the non-equilibrium spin polarization of TSSs around Fermi surface is constructed by the sum of the term $\sum_{k>0}  C^{\dagger}_{k+q,\uparrow}C_{k,\downarrow}$, which gives the clockwise SHM, and the term $\sum_{k<0} C^{\dagger}_{k,\downarrow}C_{k-q,\uparrow}$ which gives counter-clockwise SHM. Therefore the total spin polarization is along $z$ direction shown in Fig.~\ref{Spinconfig}b.

Now, let us focus on the spin polarization along $x$ direction which is completely decoupled from other spin components. Its non-purturbative dynamic equation determining
its complex frequency takes the form
\begin{eqnarray}\label{def-20}
%\left(
\left(1-\hat{D}_{xx}\right)\rho_x=0 , %\rho_x=0,
\end{eqnarray}
%\begin{eqnarray}\label{def-20}
%&&\left( \begin{array}{cc}1-(\frac{\sqrt{\Tilde{\Omega} ^2+\tilde{\Delta} _x^2} \sqrt{\Tilde{\Omega} ^2+\tilde{\Delta} _x^2+\Tilde{\Omega} _{\text{so}}^2}}{\tilde{\Delta} _x^2 \sqrt{\Tilde{\Omega} ^2+\Tilde{\Omega} _{\text{so}}^2}}-\frac{\Tilde{\Omega} }{\tilde{\Delta} _x^2}) & 0 \\ 0  &1- \frac{\sqrt{\Tilde{\Omega} ^2+\tilde{\Delta} _x^2}}{\sqrt{\Tilde{\Omega} ^2+\Tilde{\Omega} _{\text{so}}^2} \sqrt{\Tilde{\Omega} ^2+\tilde{\Delta} _x^2+\Tilde{\Omega} _{\text{so}}^2}}  \end{array}\right)\left( \begin{array}{c} \rho_x\\ \rho_z \end{array}\right)=0
%\end{eqnarray}
where
 \begin{eqnarray}\label{Dxx}
 \hat{D}_{xx}=\frac{\sqrt{\Tilde{\Omega} ^2+\tilde{\Delta} _x^2} \sqrt{\Tilde{\Omega} ^2+\tilde{\Delta} _x^2+\Tilde{\Omega} _{\text{so}}^2}}{\tilde{\Delta} _x^2 \sqrt{\Tilde{\Omega} ^2+\Tilde{\Omega} _{\text{so}}^2}}-\frac{\Tilde{\Omega} }{\tilde{\Delta} _x^2},
 \end{eqnarray}
 $l$ is the mean free path, $\Tilde{\Omega}=1-i\omega\tau_{\rm p}$, $\Tilde{\Omega}_{\rm so}=2vk_{\rm f}\tau_{\rm p}$ and $\tilde{\Delta}_{x(y)}=lq_{x(y)}$. Here we have Fourier transformed $\partial_t$ and $\partial_{x(y)}$ to $-i\omega$ and $iq_{x(y)}$ which are the frequency and wavelength of the spin polarized wave.
The in-plane spin polarization of the TSSs has a very unique property: it is proportional to the charge current perpendicular to it on the TI surface. Therefore the charge current relaxation time must be equal to the in-plane spin relaxation time.
To keep this argument in mind, we study the spin polarization along $x$-direction. First we assume that $|\Tilde{\Omega}| \ll \Tilde{\Omega}_{\rm so}$ and $\tilde{\Delta}_x \ll \Tilde{\Omega}_{\rm so}$.
The eigen-frequency of the spin dynamics takes the form
\begin{eqnarray}\label{def-18}
i\omega\tau_{\rm p}=\frac{1+\tilde{\Delta}_x^2}{2},
\end{eqnarray}
which can be Fourier transform to the real space as
\begin{eqnarray}\label{def-29}
\partial_t \rho_x =-D_{\rm s} \partial_x^2 \rho_x-\frac{\rho_x}{2\tau_{\rm p}},
\end{eqnarray}
where $D_{\rm s}=v^2_{\rm f}\tau_{\rm p}/2$.
The prior theory of spin dynamics of the TSSs (Ref.\onlinecite{Burkov:2010_a}) is based on the Kubo formula and needs to expand the spin-charge kinetic equation in terms of $\omega\tau_{\rm p}$ and $\tilde{\Delta}_x$ to their leading order. Making the same approximation, our dynamic equation of the spin polarization along $x$ direction becomes \cite{SM_T}
\begin{eqnarray}
i\omega\tau_{\rm p}=1+\frac{\tilde{\Delta}_x^2}{4}=0,
\end{eqnarray}
which is equivalent to
\begin{eqnarray}\label{def-22}
\partial_t \rho_x=-\frac{1}{2}D_{\rm s} \partial_x^2 \rho_x-\frac{\rho_x}{\tau_{\rm p}}.
\end{eqnarray}
 Eq. (\ref{def-22}) is consistent with the spin diffusion equation in Ref.~\onlinecite{Burkov:2010_a}. However we have found that when $\omega=0$, Eq. (\ref{def-22}) gives $\tilde{\Delta}_x=q^2l^2=-4$ which is of the same order but four times larger than the exact solution $\tilde{\Delta}_x^2=-1$. Both $\tilde{\Delta}_x=-4$ or $\tilde{\Delta}_x=-1$ are inconsistent with the assumption that $|\tilde{\Delta}_x| \ll 1$.
 On the other hand, the non-purtabative result of Eq.~(\ref{def-29}) shows that the spin relaxation time is $\tau_{\rm s}=2\tau_{\rm p}$, instead of $\tau_{\rm s}=\tau_{\rm p}$ indicated by Eq.~(\ref{def-22}) based on the purtabative calculation where $\tau_{\rm s}=1/|\rm{Im}(\omega)|$ is the spin life time. To better understand this result, let us compare it with the traditional DP mechanism where spin life time is inverse proportional to $\tau_{\rm p}$. We assume an initial electron state has momentum state $\psi_{\rm i,p}=\exp(ik_{\rm f}x)$ and spin state $\psi_{\rm i,s}=(1,i)^{\rm{T}}/\sqrt{2}$ on the Fermi surface. The disorder potential scatters the momentum which can be considered as rotating the effective magnetic field due to SOC with speed on the order of $1/\tau_{\rm p}$. For the traditional DP mechanism, $\Omega_{\rm so}\tau_{\rm p} \ll 1$ indicates the spin precesses too slow to follow the rotation of the SOC field. In the limit of $\tau_{\rm p}=0$ the spin have no time to precess and the spin life time is approaching infinite. This is called motional narrowing and is the unique property of the DP in the weak SOC regime. However in the strong SOC regime, $\Omega_{\rm so}\tau_{\rm p} \gg 1$ the rotation can be considered as an adiabatic process so that after disorder scattering, the momentum of the final state is $\psi_{\rm f,p}=\exp(ik_{\rm f} x\cos\theta+ik_{\rm f} y\sin\theta )$ and spin will be adiabatically rotated to the eigenstate $\psi_{\rm f,s}=(1,\exp(i\theta+\pi/2))^{\rm{T}}/\sqrt{2}$ where $\theta\neq 0$. The initial and final momentum states are orthogonal to each other indicates that the momentum memory is completely lost after the disorder scattering. However, the initial and final spin states has finite overlap, $|\langle \psi_{\rm f,s}|\psi_{\rm i,s}\rangle|^2=\cos^2\theta$, whose average value is $1/2$ and consistent to our conclusion that the spin life time is two times longer than the momentum life time. On the other hand,
 it is well known that charge transport time $\tau_{\rm{tr}}$ for steady current in the system with Dirac-like Hamiltonian
 is double of the momentum scattering time $\tau_{\rm p}$ due to the spin-momentum locking \cite{Culcer:2010_a}.
 Therefore, our non-perturbative theory systemically proves that the spin relaxation time and charge transport time $\tau_{\rm{tr}}$
 are equal for the TSSs, $\tau_{\rm s}=\tau_{\rm{tr}}$ which is qualitatively different to the traditional DP mechanism and consistent with one of the unique properties of the Dirac-like Hamiltonian, that the spin operator is proportional to the velocity operator.

In the case of uniform in-plane spin polarization, without assuming $|\tilde{\Omega}| \ll \tilde{\Omega}_{\rm so}$, there is another solution of Eq.~\ref{def-20}
\begin{eqnarray}\label{def-23}
i\omega\tau_{\rm p}= \frac{3}{4}\pm i\tilde{\Omega}_{\rm so},
\end{eqnarray}
which gives a damped oscillatory spin dynamic mode. Due to the strong SOC, this spin oscillation is accompanied by an AC current with the same frequency. This AC current may provide another way to detect the transport property of TSSs.

Now, we focus on the spin polarization perpendicular to the surface. Based on Eq. (\ref{GF-12},\ref{GF-13}), the dynamic equation of the spin polarization along z direction takes the form
\begin{eqnarray}\label{def-1}
1-\hat{D}_{zz}=0,
 \end{eqnarray}
 where \begin{eqnarray}\label{Dzz}
 \hat{D}_{zz}=\frac{\sqrt{\tilde{\Omega} ^2+\tilde{\Delta} _x^2}}{\sqrt{\tilde{\Omega} ^2+\tilde{\Omega} _{\text{so}}^2} \sqrt{\tilde{\Omega} ^2+\tilde{\Delta} _x^2+\tilde{\Omega} _{\text{so}}^2}}.
 \end{eqnarray}
When $\omega_z=0$ for the steady spin polarization, we have
\begin{eqnarray}\label{def-17}
\tilde{\Delta}_x=i\sqrt{2+\tilde{\Omega}_{\rm so}^2}\approx 2ik_{\rm f} l,
\end{eqnarray}
which means the diffusive length of out of plane spin is of the same order as the Fermi wave length. This is in contrast to the diffusive length in other materials such as the GaAs/AlGaAs where it is of the same order of (or much smaller) then the mean free path $l$ in the strong (weak) SOC regime \cite{Bernevig:2008_a, LiuXin:2011_a, Burkov:2004_a, Stanescu:2007_a}.
\begin{figure}
\centering
\begin{tabular}{l}
\includegraphics[width=0.8\columnwidth]{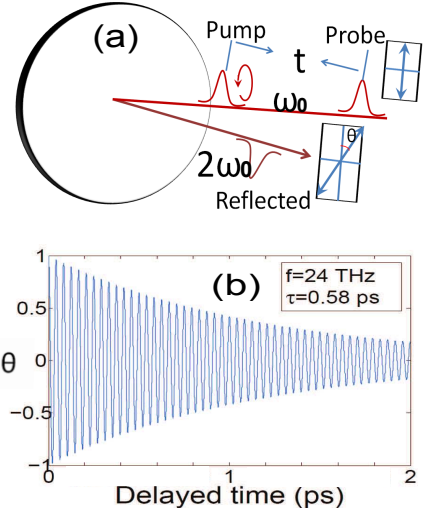}
\end{tabular}
\caption{(a) The second harmonic pumping probe measurement. The circularized pump light with the frequency $\omega_0$ is followed by the linear polarized probe light with the same frequency with the time delay $t$. The reflected light with $2\omega_0$ frequency has a $\theta$ rotated angle of the linear polarization due to the Kerr effect. (b) The evolution of the uniform out of plane spin polarization. According to our theory, we take the experimental parameters from the sample Q2 in table 1 of the Ref.\cite{Qu:2010_a} as $k_{\rm f}=0.032{\AA}^{-1}$, $E_{\rm f}=84 \rm{meV}$ and $k_{\rm f} l=69$.}
\label{S-TI}
\end{figure}
When the spin is uniformly polarized perpendicular to the surface, the eigenfrequency has the form
\begin{eqnarray}\label{def-15}
 i\omega\tau_{\rm p} =\frac{1}{2}(1-\sqrt{1-4\tilde{\Omega}_{\rm so}^2})\approx \frac{1}{2} \pm i\tilde{\Omega}_{\rm so},
\end{eqnarray}
where the approximation is valid when $\tilde{\Omega}_{\rm so} \gg 1$.  Eq.~(\ref{def-15}) indicates that the spin along the $z$-direction will
oscillate in a damped form with the frequency equal to $\tilde{\Omega}_{\rm so}=2 vk_{\rm f} \tau_{\rm p}$ and the relaxation time given as $\tau_{\rm s}=\tau_{\rm{tr}}=2\tau_{\rm p}$, which is the same to the decay rate of the in-plane spin polarization. We notice that although the nonuniform spin dynamics on the TI surface are very different to that in 2DEG with strong SOC, the uniform out-of-plane spin dynamics follow the same dynamic equation. Experiments\cite{Leyland:2007_a} have observed $\tau_{\rm s}=2\tau_{\rm p}$ in the 2DEG with strong SOC which is independent on the Fermi energy, the strength of SOC and temperature. Therefore, we expect the similar results for the spin dynamics on the TI surface and therefore provide a reliable way to measure the charge transport time $\tau_{\rm tr}$ of TSSs. Because the spin along $z$-direction is not coupled to any other spin and charge density, this oscillation will not induce current oscillation or charge density oscillation. The experiments have been able to freely tune the Fermi surface of TI close to or away from the Dirac point in a wide range \cite{Hsieh:2009_a}. A circular polarized pump beam can generate spin non-equilibrium polarization of TSSs which has been predicted theoretically \cite{Lu:2010_a} and verified experimentally \cite{LiSimian:2011_a}. The spin dynamics of the TSSs can be detected by the time-resolved second harmonic optical pump-probe measurements. As only TSSs contribute the second harmonic
generation\cite{Hsieh:2011_a,Hsieh:2011_b}, this type of measurement can isolate the surface response from the bulk. If the Fermi surface is turned to $84$ meV \cite{Qu:2010_a} above the Dirac point, the experiment should observe about $24$ THz oscillation through the Kerr rotation angle. The femtosecond pump probe spectroscopy can be used to measure this oscillation.  At the same time, the decay rate of this spin oscillation is $\tau_{\rm{tr}}$ which gives us the charge transport time of the TSSs (Fig.\ref{S-TI}).

In conclusion, we develop a non-purtabative method to study the generalized spin dynamic of the TSSs which is shown to be qualitatively different to the traditional DP mechanism. We found the in-plane spin relaxation time of the pure exponential decay mode is exactly equal to the charge transport time. We also predict two fast oscillatory modes of spin polarization perpendicular and parallel to the surface of a topological insulator, which cannot be obtained from the prior spin diffusion equations, based on perturbative approaches. At last, we shown how to read the charge transport properties of TSSs from the out of plane spin dynamics.

We acknowledge Chia-Ren Hu, Artem Abanov, Cenke Xu, Jainendra Jain and Chao-Xing Liu for very helpful discussion and support from DMR-1105512, NHARP, and ONR-000141110780. X.L. acknowledges partial support by the DOE under Grant No. DE-SC0005042.

%%%%%%%%%%%%%%%%%%%%%%%%%%%%%%%%%%%%%%%%%%%%%%%%%%%%%%%%%%%%%%%%%%%%%%%%%%%%%%%%%%%%%%%%%%%%%%%%%%
%\bibliography{spin-charge,TI_references}
%

%%%%%%%%%%%%%%%%%%%%%%%%%%%%%%%%%%%%%%%%%%%%%%%%%%%%%%%%%%%%%%%%%%%%%%%%%%%%%%%%%%%%%%%%%%%%%%%%%%%

\noindent

%%%%%%%%%%%%%%%%%%%%%%%%%%%%%%%%%%%%%%%%%%%%%%%%%%%%%%%%%%%%%%%%%%%%%%%%%%%%%%%%%%%%%%%%%%%%%%%%%%%%%%%%%%%%%%%%%%%%%%%%%%%%

\end{document}